\RequirePackage{lineno}
\documentclass[twocolumn,showpacs,aps,prd,superscriptaddress,floatfix]{revtex4-1}
\usepackage{graphicx}  
\usepackage{dcolumn}   
\usepackage{bm}        
\usepackage{amssymb}   
\usepackage{amsfonts}  
\usepackage{hyperref}
\usepackage{amsmath}
\usepackage{mathtools}
\usepackage{xspace}

\def\gev    {\ensuremath{\mathrm{GeV}}\xspace}

\def\gevcc  {\ensuremath{\mathrm{GeV}/c^2}\xspace}
\def\mev    {\ensuremath{\mathrm{MeV}}\xspace}

\def\numjpsi   {(1310.6 \pm 7.0)\times 10^6}
\def\numRomega {8.1\times 10^{-5}}
\def\numRphi   {3.4\times 10^{-4}}
\def\numBomega {7.3\times 10^{-5}}
\def\numBphi   {1.7\times 10^{-4}}
\def\egamext  {E^{\rm Extra}_{\gamma}}
\def\mvrecoil {M_{\rm recoil}^{V}}
\def\etot     {E^{\rm tot}}
\def\ptot     {p^{\rm tot}}
\def\Nsig     {N_{\rm sig}}
\def\piz      {\pi^0}
\def\ee       {e^+e^-}
\def\jpsi     {J/\psi}
\def\ppp      {\pi^+\pi^-\pi^0}
\def\kk       {K^+K^-}
\def\vomegaphi {V=\omega, \phi}
\def\jpsiveta {J/\psi\to V \eta}
\def\jpsiomeeta{J/\psi\to\omega\eta}
\def\jpsiphieta{J/\psi\to\phi\eta}
\def\omegavis {\omega\to\pi^+\pi^-\pi^0}
\def\omegainv {\omega\to \rm {invisible}}
\def\phivis   {\phi\to K^+K^-}
\def\phiinv   {\phi\to \rm {invisible}}
\def\etappp   {\eta\to \pi^+\pi^-\pi^0}
\def\Romega   {\frac{\mathcal{B}(\omega \rightarrow \rm{invisible})}{\mathcal{B}(\omega \rightarrow \pi^+\pi^-\pi^0)}}
\def\Rphi     {\frac{\mathcal{B}(\phi \rightarrow \rm{invisible})}{\mathcal{B}(\phi \rightarrow K^+K^-)}}
\def\Bomega   {\mathcal{B}(\omega \rightarrow \rm{invisible})}
\def\Bphi     {\mathcal{B}(\phi \rightarrow \rm{invisible})}

\def\simge{\mathrel{
   \rlap{\raise 0.511ex \hbox{$>$}}{\lower 0.511ex \hbox{$\sim$}}}}
\def\simle{\mathrel{
   \rlap{\raise 0.511ex \hbox{$<$}}{\lower 0.511ex \hbox{$\sim$}}}}

\begin{document}
\title{\bf Search for invisible decays of \boldmath{$\omega$} and \boldmath{$\phi$}  with  \boldmath{$J/\psi$} data at BESIII}

\author{\small
M.~Ablikim$^{1}$, M.~N.~Achasov$^{10,d}$, S. ~Ahmed$^{15}$, M.~Albrecht$^{4}$, M.~Alekseev$^{55A,55C}$, A.~Amoroso$^{55A,55C}$, F.~F.~An$^{1}$, Q.~An$^{52,42}$, Y.~Bai$^{41}$, O.~Bakina$^{27}$
% ...
\\
\vspace{0.2cm}
(BESIII Collaboration)\\
\vspace{0.2cm} {\it
$^{1}$ Institute of High Energy Physics, Beijing 100049, People's Republic of China\\
% ...
$^{4}$ Bochum Ruhr-University, D-44780 Bochum, Germany\\
% ...
$^{10}$ G.I. Budker Institute of Nuclear Physics SB RAS (BINP), Novosibirsk 630090, Russia\\
% ...
$^{15}$ Helmholtz Institute Mainz, Johann-Joachim-Becher-Weg 45, D-55099 Mainz, Germany\\
% ...
$^{27}$ Joint Institute for Nuclear Research, 141980 Dubna, Moscow region, Russia\\
% ...
$^{41}$ Southeast University, Nanjing 211100, People's Republic of China\\
$^{42}$ State Key Laboratory of Particle Detection and Electronics, Beijing 100049, Hefei 230026, People's Republic of China\\
% ...
$^{52}$ University of Science and Technology of China, Hefei 230026, People's Republic of China\\
% ...
$^{55}$ (A)University of Turin, I-10125, Turin, Italy; (B)University of Eastern Piedmont, I-15121, Alessandria, Italy; (C)INFN, I-10125, Turin, Italy\\
% ...
$^{d}$ Also at the Novosibirsk State University, Novosibirsk, 630090, Russia\\
% ...
}\vspace{0.4cm}}

\begin{abstract}
  Using a data sample of $\numjpsi$ $\jpsi$ events collected with the BESIII detector operating at the BEPCII collider,
  we perform the first experimental search for invisible decays of a
  light vector meson ($\vomegaphi$) via $\jpsiveta$ decays. The decay
  of $\etappp$ is utilized to tag the $V$ meson decaying into the
  invisible final state.   No evidence for a significant  invisible
  signal is observed, and the upper limits on  the ratio of branching
  fractions at the 90\% confidence level   are determined to be $\Romega<\numRomega$ and $\Rphi< \numRphi$.
  By using the world average values of $\mathcal{B}(\omegavis)$ and $\mathcal{B}(\phivis)$, the upper limits on the decay branching fractions at the 90\% confidence level are set as  $\Bomega< \numBomega$ and $\Bphi< \numBphi$, respectively.
\end{abstract}

\pacs{14.40.Be, 95.35.+d}

\maketitle

\section{Introduction}
  Although there is strong evidence from many astrophysical observations for the existence of dark matter, its nature is still mysterious. Dark matter is invisible in the entire electromagnetic spectrum, and its existence is inferred via  gravitational effects only. Any information about its interactions with a Standard Model (SM) particle would shed light on the nature of dark matter. Quarkonium states, whose constituents are a quark and its own anti-quark, are expected to annihilate into a neutrino-pair ($\nu \overline{\nu}$) via a  virtual $Z^0$ boson. However, the process is very rare in the SM~\cite{chang}. The branching fraction of the invisible decays might be enhanced by several orders of magnitude in the presence of  light dark matter (LDM) particles  $\chi$~\cite{ldm,fayet,meVDM} as described in Refs.~\cite{MCelarth1, MCelarth2}.

  The LDM particles, which are in the kinematic reach of BESIII, may provide one possible explanation of the feature of the  511 keV gamma ray excess from the galactic center observed by the INTEGRAL satellite~\cite{ingral}. The smooth symmetric morphology of 511 keV gamma emission is believed to originate from the annihilation of LDM particles into  $\ee$ pairs~\cite{ldm,ldm1}. The LDM particles  can have adequate relic abundance to account for the nonbaryonic dark matter~\cite{nonbaryonic} in the universe, if they couple with the SM particles via a new light gauge boson $U$~\cite{DM} or the exchange of  heavy fermions in the case of scalar dark matter~\cite{ldm,fayet}.  One of the most popular LDM candidates is the neutralino predicted by the Next-to-Minimal Supersymmetric Standard Model ~\cite{nmssm}, which is stable due to the conserved R-parity~\cite{Rparity}.

  The BESII~\cite{besinv} and BaBar~\cite{babarinv} experiments have set  the most stringent upper limits on the invisible decays of $\jpsi$ and $\Upsilon(1S)$, respectively, which are still above the SM predictions~\cite{chang}.  The experimental exploration of invisible decays for other quarkonium states ($q \overline{q},~ q=u,d$ or $s$) may help to constrain the masses of the LDM particles  and the coupling of the $U$ boson to  light quarks~\cite{Nicolas, invisibleuds}. The branching fraction  $\mathcal{B}(V \to \chi\chi)$  ($\vomegaphi$) is predicted to be up to the level of $10^{-8}$ by assuming the same cross section for the time reversed processes, $\sigma(q\overline{q} \to \chi\chi) \simeq \sigma(\chi \chi \to q\overline{q})$~\cite{MCelarth2}. The search for these decays can be performed via a two-body decay process of $\jpsiveta$.
  In this paper, we report  the first experimental search for the invisible decays of $\omega$ and $\phi$ mesons via $\jpsiveta$  using $\numjpsi$ $\jpsi$ events collected with the BESIII detector in 2009 and 2012~\cite{njps}.

\section{The BESIII experiment and Monte Carlo simulation}
BESIII is a cylindrical particle physics detector located  at the BEPCII facility, a double-ring $\ee$ collider with a peak luminosity of $10^{33}$~cm$^{-2}$s$^{-1}$ at the center-of-mass (CM) energy of 3.773~$\gev$. It has four detector sub-components with  a coverage of 93\% of the total solid angle as described in Ref.~\cite{bes3_nim}. Charged particle momenta are measured in a 43-layer helium based main drift chamber (MDC) operating with  a 1.0~T (0.9~T) solenoidal magnetic field during 2009 (2012) $\jpsi$ runs. Charged particle identification (PID) is performed using the energy loss (d$E$/d$x$) measured in the MDC with a resolution better than 6\%, and a time-of-flight (TOF) system consisting of 5 cm thick plastic scintillators with a time resolution of 80 ps in the barrel region and 110 ps in the end-cap region, respectively. Photon and electron energies are measured in a CsI(Tl) electromagnetic calorimeter (EMC). The energy (position) resolution of the EMC for 1 GeV electrons and photons is 2.5\% (6~mm) in the barrel and 5.0\% (9~mm) in the end-cap regions. The muons are identified in a muon counter (MUC) containing nine (eight) layers of resistive plate chamber counters interleaved with steel in the barrel (end-caps) region. The MUC provides a spatial resolution better than 2 cm. 

  A large number of Monte Carlo (MC) events are produced to optimize the event selection criteria, to study the potential backgrounds and to determine the reconstruction efficiencies. The MC simulation includes the detector response and signal digitization models simulated by {\sc Geant4}~\cite{geant4} and takes into account time-dependent detector effects, such as beam related backgrounds and detector running conditions during the data-taking period. An MC sample of $1225\times 10^6$  inclusive $\jpsi$ events is generated for background studies. The known $\jpsi$ decay modes are generated by the {\sc EvtGen} generator package~\cite{evtgen} with the branching fractions taken from the Particle Data Group (PDG)~\cite{pdg}, while the remaining unknown $\jpsi$ decay modes are generated by the {\sc LUNDCHARM}~\cite{lundcharm} generator. The production of the $\jpsi$ resonance via $\ee$ annihilation is simulated by the {\sc KKMC}~\cite{kkmc} including the effects of the beam energy spread and initial state radiation (ISR).  We use a helicity amplitude model for the $J/\psi \to V \eta$ decay, an $\omega$ Dalitz plot distribution model for the $\omega \to \ppp$ decay~\cite{omegadalitz}, an $\eta$ Dalitz plot distribution model for the $\eta \to \ppp$ decay~\cite{etadalitz}, a vector meson decaying to a pair of scalar particles model for the $\phi \to K^+K^-$ decay, and a phase space model for $V \to \nu \overline{\nu}$ decays~\cite{evtgen}.

\section{Analysis Strategy}
\label{strategy}
  The search for invisible decays of $\omega$ and $\phi$ mesons is performed by using the two-body $\jpsiveta$ decay process. The candidate events are tagged with the $\eta$ reconstructed from its $\ppp$ decay mode, and the mass distribution of the system recoiling against the $\eta$ candidate is used to investigate invisible decays of $\omega$ and $\phi$ mesons. The more prominent decay mode of $\eta\to\gamma\gamma$ is not used for the tagging due to the huge background contamination.

  In order to minimize the systematic uncertainty, the  decays  $\omegavis$ and $\phivis$ from $\jpsiveta$ decays are reconstructed as  reference channels. The ratio of the branching fraction of the invisible decay to that of the visible decay of V mesons is measured by
\begin{equation}
\frac{\mathcal{B}(V\rightarrow \rm invisible)}{\mathcal{B}(V\rightarrow \rm visible)} = \frac{N_{\rm sig}^{\rm invisible}\cdot \epsilon^{\rm visible}}{N_{\rm sig}^{\rm visible}\cdot \epsilon^{\rm invisible}},
\label{bfrat}
\end{equation}
\noindent
  where $N_{\rm sig}^{\rm invisible}$ and $N_{\rm sig}^{\rm visible}$ are the numbers of signal events for the invisible and visible decays, respectively, the $\epsilon^{\rm invisible}$ and $\epsilon^{\rm visible}$ are the corresponding detection efficiencies. By applying this method, the systematic uncertainties associated with  the total number of $\jpsi$ events, the branching fractions $\mathcal{B}(\jpsiveta)$ and $\mathcal{B}(\etappp)$, and the reconstruction of $\eta$ candidates (such as tracking, PID and photon detection efficiency \emph{etc}.) are canceled.

\section{Data analysis}
The charged tracks are measured in the MDC with the polar angle $\theta$ satisfying $|\cos\theta|< 0.93$. They must have the points of closest approach to the beam line within $\pm 10.0\;\mathrm{cm}$ from the interaction point along the beam direction and $\pm 1.0\;\mathrm{cm}$ in the plane perpendicular to the beam. PID 
for the charged tracks is accomplished by combining the measured energy loss (d$E$/d$x$) in the MDC, the flight time obtained from the TOF and the electromagnetic cluster shower information from the EMC to form the likelihoods for electron, kaon and pion hypotheses. A charged pion is identified by requiring  the PID probability  of  its  pion  hypothesis  to  be  larger  than  the  kaon  and  electron hypotheses.

  The photon candidates, reconstructed using the clusters of energy deposited in the EMC, are selected with a minimum energy of $25~\mev$ in the barrel region ($|\cos\theta| < 0.8$) or $50~\mev$ in the end-cap region ($0.86 < |\cos\theta| < 0.92$). To improve the reconstruction efficiency and energy resolution, the energy deposited in the nearby TOF counters is included. The angle between a photon and the nearest extrapolated track in the EMC is required to be greater than $10$ degrees to avoid any overlap between charged  and neutral tracks.  In order to suppress  electronic noise and energy deposits unrelated  to the signal events, the EMC timing of the photon candidate is required to be within 700~ns relative to the event start time. A $\piz$ candidate is reconstructed from a photon pair candidate, and the two-photon invariant mass  is constrained  to the nominal value of the $\piz$ meson~\cite{pdg} by performing a kinematic fit.

\subsection{The invisible decays of $\omega$ and $\phi$ mesons}
\label{visibledecayinv}

 For studies of the invisible decays of a $V$ meson using the decay chain $\jpsiveta$, $\etappp$, the event candidate is required to have two oppositely charged tracks  identified as pions. A vertex fit is performed to these two charged tracks to insure that they originate from a common vertex.  The $\piz$ candidate for which the  $\ppp$  invariant mass ($M_{\ppp}$) is closest to the nominal mass of the $\eta$ meson~\cite{pdg} is considered as originating from  the $\eta$ decay. An $\eta$ candidate is required to have $M_{\ppp}$ within [0.52,0.57]~$\gevcc$. With the above requirements, MC studies indicate that the dominant backgrounds are from $\jpsiveta$ with the $V$ meson decaying into purely neutral final states, such as  $\omega \rightarrow \gamma \piz$ and $\phi \rightarrow K_S K_L$, $K_S \rightarrow \piz \piz$. Thus, $\egamext$  is required to be less than 0.2~$\gev$, where $\egamext$ is the sum of energies of the extra photons, which are not used in the $\eta$ reconstruction. Furthermore, the polar angle of the system recoiling against the selected $\eta$ candidate, $\theta_{\rm recoil}$, is required to satisfy $|\cos\theta_{\rm recoil}| < 0.7$ to further eliminate the background contributions from  $\jpsi\to X \eta$, where X can be any final state emitted in the region which is not covered by the acceptance of the detector.

\begin{figure}[htbp]
\begin{center}
\includegraphics[width=0.4\textwidth]{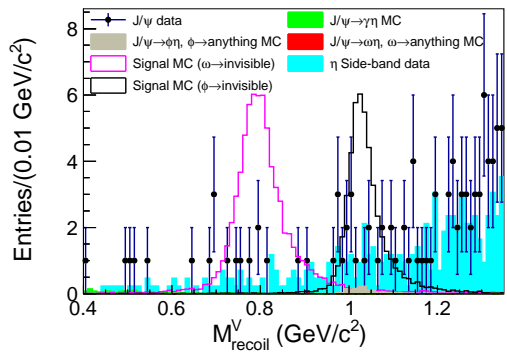}
\caption{(Color online) Invariant mass recoiling against the selected $\eta$ candidate ($\mvrecoil$) for data (black dot points with error bars), signal MC samples (pink and black histograms for $\omega$ and $\phi$, respectively) and various expected backgrounds shown as different colored histograms.}
\label{fig:mrecoil}
\end{center}
\end{figure}

  The signals of the invisible decays of $\omega$ and $\phi$ mesons are inferred from  the invariant mass of the system recoiling against the selected $\eta$ candidate, defined as $\mvrecoil\equiv\sqrt{(E_{cm}-E_{\ppp})^2-P_{\ppp}^2}$, where $E_{cm}$ is the CM energy, and $E_{\ppp}$ and $P_{\ppp}$ are the energy and momentum of the $\ppp$ system in the CM frame, respectively. The $\mvrecoil$ distribution of the  event candidates for the data range [0.40, 1.35]~$\gevcc$ is shown in Fig.~\ref{fig:mrecoil}. The expected distributions for $\omega$ and $\phi$ invisible decay signals by MC simulation are also depicted in the plot. Detailed studies of the inclusive $\jpsi$ decay sample indicate that the non-peaking backgrounds are dominated by  processes with non-$\eta$ mesons in the final state, which can be evaluated with the normalized events in the $\eta$ mass sideband regions, as shown by a cyan histogram in Fig.~\ref{fig:mrecoil}. The non-peaking background from $\jpsi \to \gamma \eta$, which has a large branching fraction, is evaluated to be $1.8$ events with negligible uncertainties by using an exclusive MC sample normalized according to the branching fractions quoted in the PDG~\cite{pdg}, and is ignored in the following analysis.
  The possible peaking background is from the decay $\jpsiveta$ with the $V$ meson decaying visibly. The numbers of peaking backgrounds are evaluated to be $0.1$ for $\jpsiomeeta$ and $2.0$ for $\jpsiphieta$  with negligible uncertainty using the simulated MC samples normalized according to the measured branching fractions of $\jpsiveta$ described in Sec.~\ref{visibledecayomega} and  \ref{visibledecayphi}, respectively, and the corresponding distributions are presented in Fig.~\ref{fig:mrecoil}. The backgrounds from other sources are  negligible. The $\mvrecoil$ distributions of simulated signal MC events for invisible decays of $\omega$ and $\phi$ mesons are observed to be well consistent with the data and MC simulations of their visible decays described in Sec.~\ref{visibledecayomega} and  \ref{visibledecayphi}, respectively.

\begin{figure}[htbp]
\begin{center}
\includegraphics[width=0.4\textwidth]{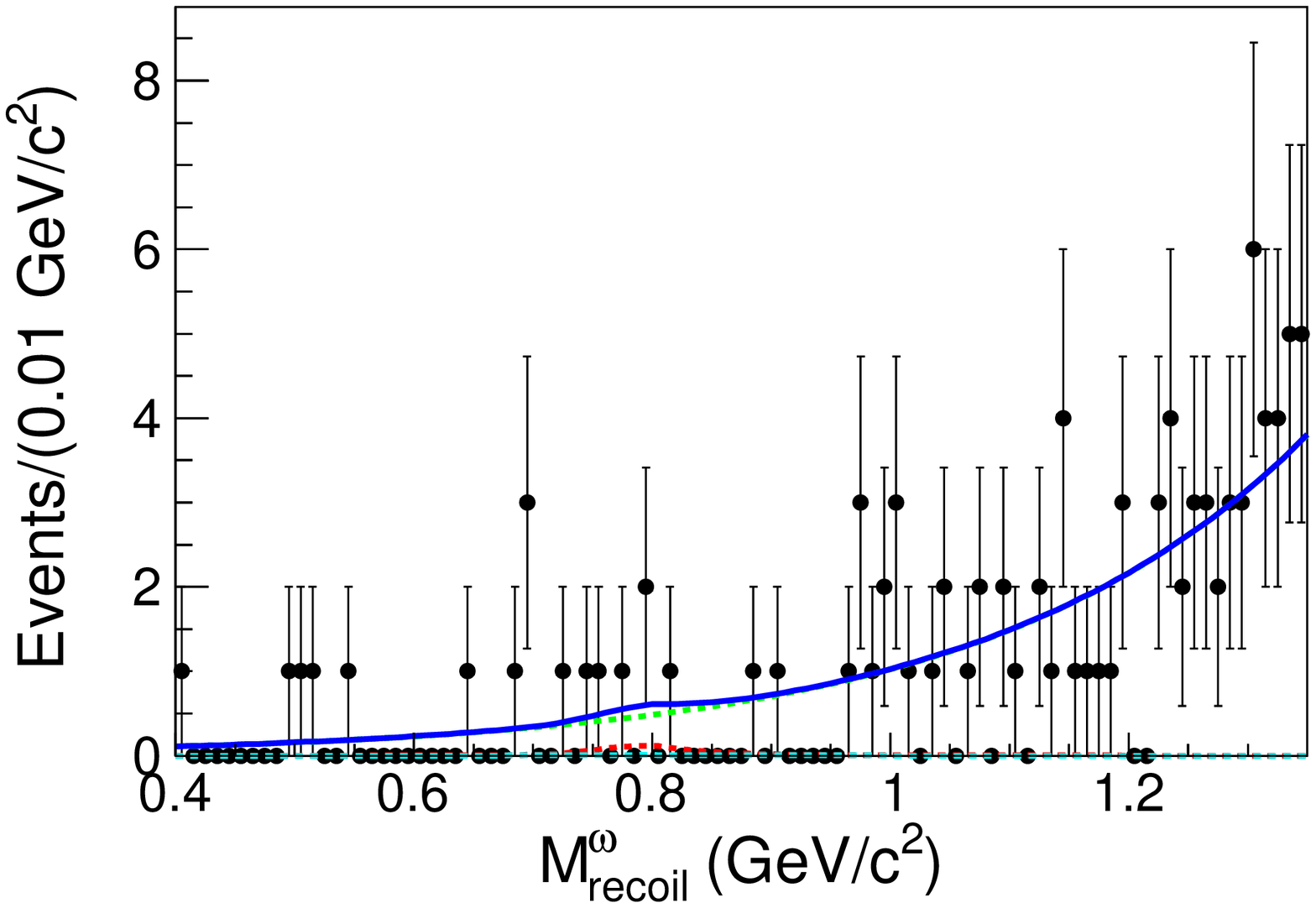}
\includegraphics[width=0.4\textwidth]{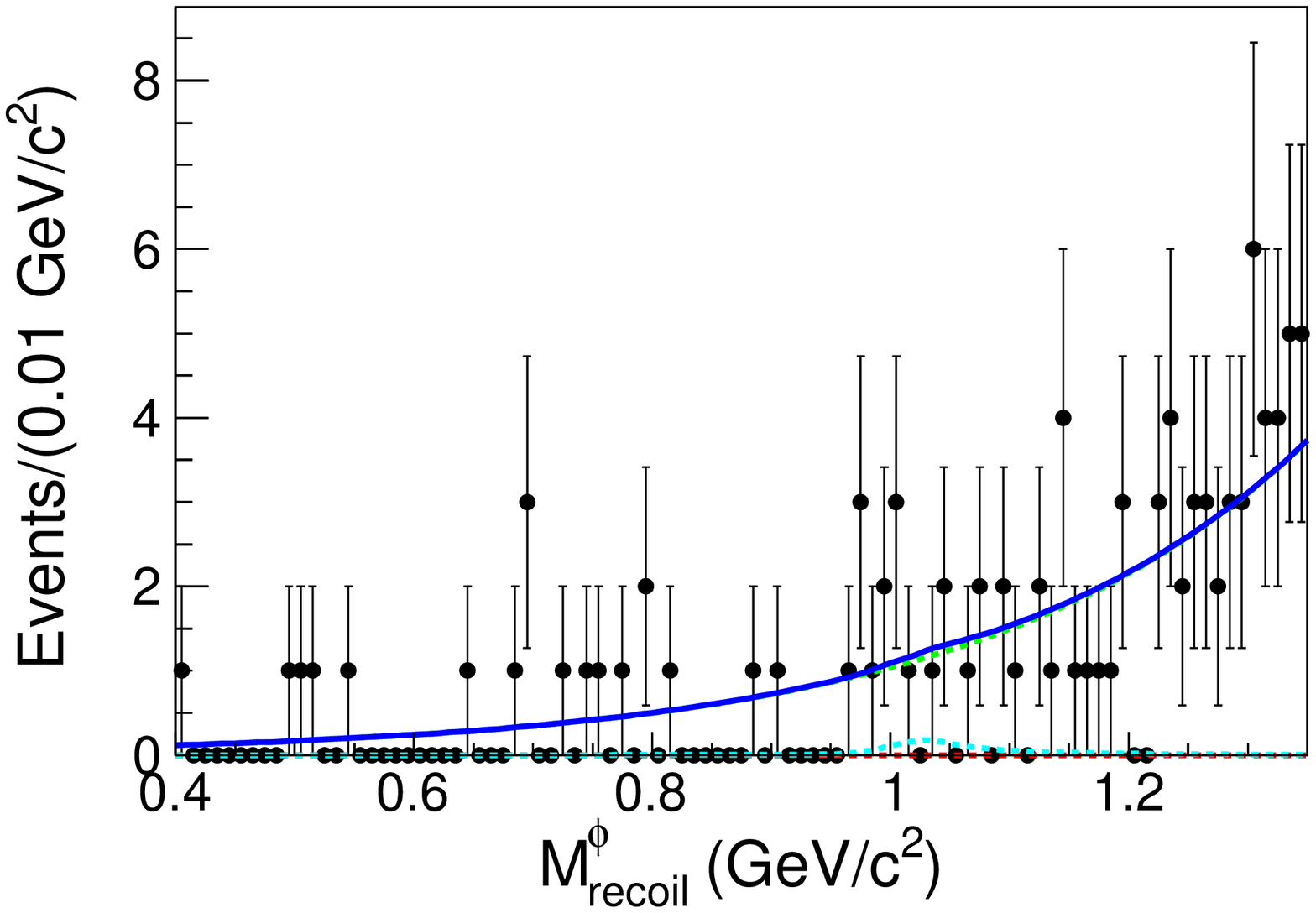}
\caption{(Color online) Fit to the $\mvrecoil$ distribution for $\omega$ (top) and $\phi$ (bottom) signals. The data are shown by the dots with error bars, the non-peaking background by the green dashed curve, the peaking background by the cyan dashed curve, the signal by the red dashed curve. and the total fit by the blue solid curve.}
\label{fig:proj}
\end{center}
\end{figure}

An extended maximum likelihood (ML) fit to the $\mvrecoil$ distribution is performed to obtain the signal yield ($\Nsig$). The probability density function (PDF) of the  $V$ meson invisible decay signal and peaking background is described by their MC simulated shapes, while that of the non-peaking background is represented by an increasing exponential function. In the fit, the number of peaking background events is fixed, while the parameters of the non-peaking background PDF and the yields for signal and non-peaking background events are free parameters in the fit. The ML fit yields $\Nsig= 1.4 \pm 3.6$ events for the $\omegainv$ decay and $\Nsig = -0.6 \pm 4.5$ for the $\phiinv$ decay, respectively. The obtained $\Nsig$ events for both decay modes are consistent with zero, and no evidence of invisible decays of $\omega$ and $\phi$ mesons is observed. The  fitted $\mvrecoil$ are shown in Fig.~\ref{fig:proj}. The corresponding signal detection efficiencies, estimated with the MC simulation, are $20.5\%$ and $21.3\%$ for $\omega$ and $\phi$ invisible decays, respectively.

\subsection{The visible decay mode {\boldmath $\omegavis$}}
\label{visibledecayomega}
\begin{figure}[htbp]
\begin{center}
 \includegraphics[width=0.45\textwidth]{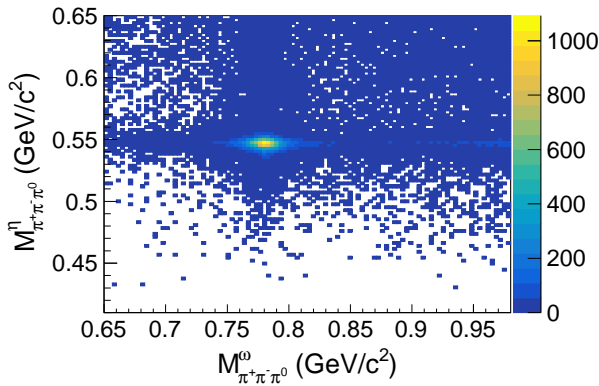}
 \caption{(Color online) Distribution of  $M_{\ppp}^{\omega}$ versus $M_{\ppp}^{\eta}$  for data.}
\label{fig:2dplotomega}
\end{center}
\end{figure}

  The candidate events of $\jpsiomeeta$ with subsequent decays $\omegavis$ and $\etappp$ are required to have four charged tracks with net charge zero and at least two independent $\piz$ candidates without sharing the same photon. The four charged tracks are assumed to be pions and required to originate from a common vertex by performing a vertex fit.
  For an event with multiple $\piz\piz$ pair candidates, the one with the least value of $\ptot$ is selected, where $\ptot$ is the total momentum of the $2(\ppp)$ candidates. The total energy ($\etot$) of the selected candidate is also required to satisfy $\etot>2.95~\gev$. For a selected $2(\ppp)$ final state, the combinations of $\ppp$ for $\omega$ and $\eta$ signals are determined by
\begin{equation}
\chi_{\omega\eta}^2= \frac{(M_{\ppp}^{\omega} -M_{\omega})^2}{\sigma_{\omega}^2}+\frac{(M_{\ppp}^{\eta} -M_{\eta})^2}{\sigma_{\eta}^2},
\end{equation}
where  $M_{\ppp}^{X}$ ($X=\omega,\eta$) is the invariant mass of the $\ppp$ combination for the $X$ candidate, $M_{X}$ is the nominal  $X$ meson mass quoted in the PDG~\cite{pdg}, and $\sigma_{X}$ is the corresponding mass resolution determined from the signal MC simulation. All eight combinations of $(\ppp)_{\omega}$ versus $(\ppp)_{\eta}$ are explored, and the one with the least $\chi^2_{\omega\eta}$ is selected. In order to improve the purity of pions in the $\eta \to \ppp$ decay and to minimize the systematic uncertainty in the analysis, PID for charged pions from the $\eta$ decay is performed, but no PID requirement for those from $\omega$ decay due to tiny expected background contribution from  $\omega \to l^+l^- \pi^0$ ($l=e,\mu$) in the full $J/\psi$ data sample. Similarly to the invisible decay, the polar angle of the system recoiling against the $\eta$ candidate $\theta_{\rm recoil}$ is required to satisfy $|\cos\theta_{\rm recoil}| < 0.7$ to minimize the systematic uncertainty.
  The selected candidate events are further required to have $M_{\ppp}^{\omega}$ and $M_{\ppp}^{\eta}$ in the ranges  [0.65, 0.98] and [0.41, 0.65]~$\gevcc$, respectively.  Figure~\ref{fig:2dplotomega} shows the two-dimensional (2D) histogram of  $M_{\ppp}^{\omega}$ versus $M_{\ppp}^{\eta}$  for data.

  The remaining backgrounds are dominated by those with the same final state  as the signal, but neither $\omega$ nor $\eta$ intermediate states included (named BKGI thereafter) or without either $\omega$ or $\eta$ intermediate state (named BKGII thereafter). In addition, there is a small peaking background for both  $M_{\ppp}^{\omega}$ and $M_{\ppp}^{\eta}$  simultaneously (named BKGIII thereafter),  dominated by $\jpsiomeeta$ with the subsequent decays  $\omegavis$ and $\eta\to\gamma\pi^+\pi^-$. Consequently, the contributions of BKGI and BKGII are determined by
performing a  2D ML fit to  $M_{\ppp}^{\omega}$ and $M_{\ppp}^{\eta}$, while  BKGIII is determined by using a corresponding
exclusive MC sample normalized according to the branching fractions
quoted in PDG~\cite{pdg}. The BKGIII yield, estimated to be  $1085.8 \pm 126.6$
events, which  uncertainty includes the uncertainties of both the total number of $J/\psi$ events and the branching fractions of the corresponding decay process, is subtracted from the signal yield obtained from the 2D ML
fit, eventually.

\begin{figure}[htbp]
  \begin{center}
\includegraphics[width=0.4\textwidth]{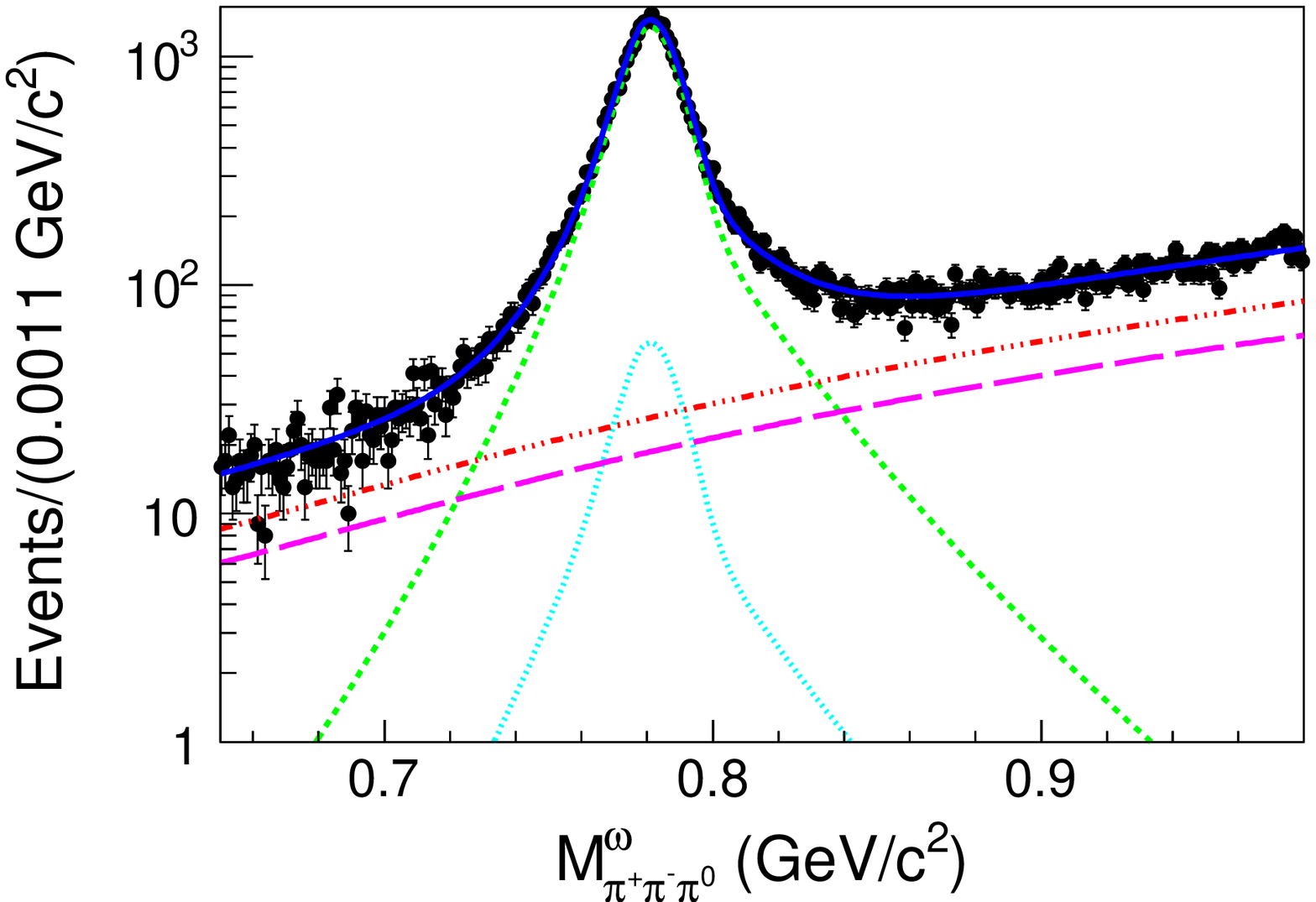}
\includegraphics[width=0.4\textwidth]{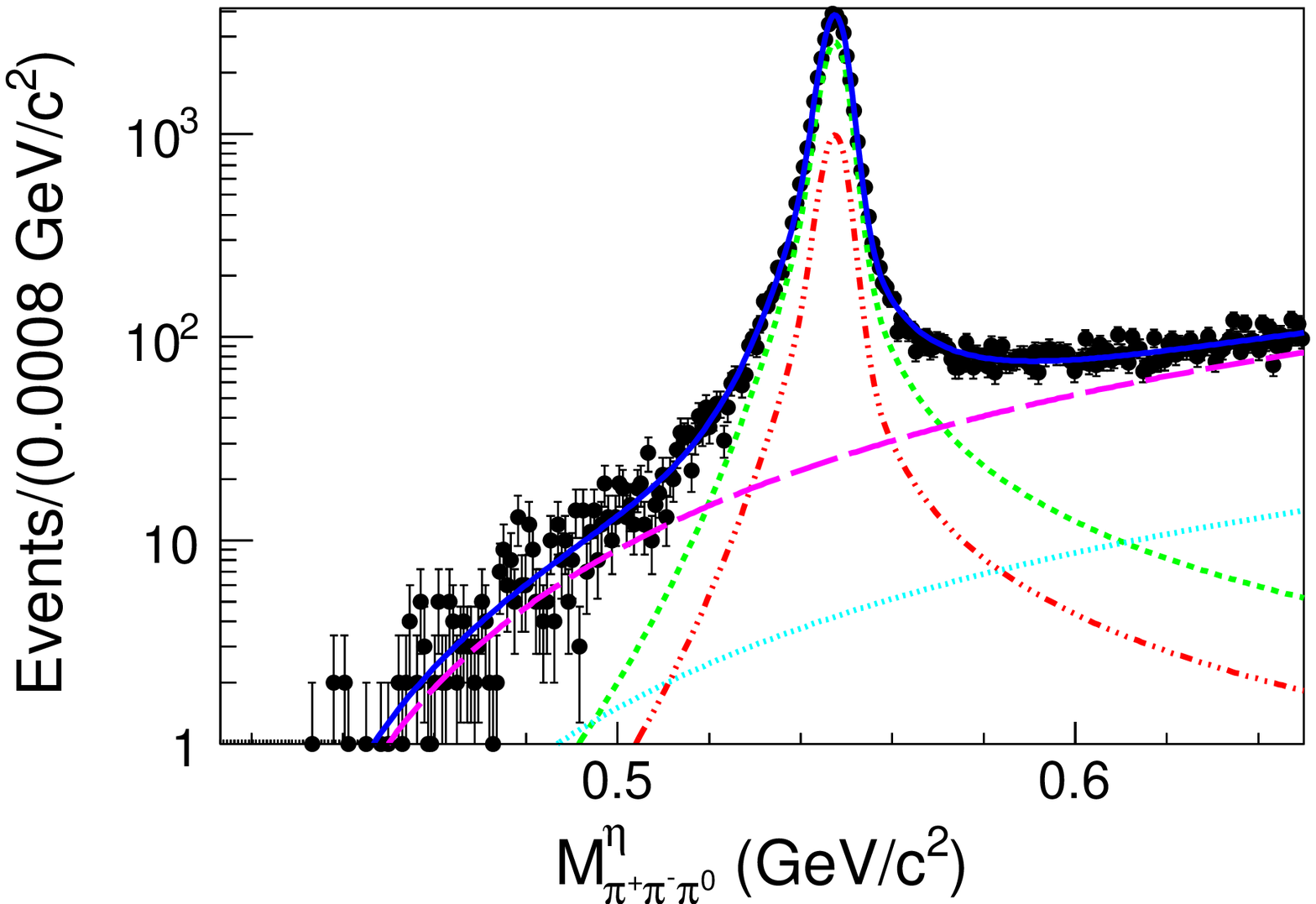}    
 \caption{(Color online) Projections of the 2D fit to the $M_{\ppp}^{\omega}$ (top) and $M_{\ppp}^{\eta}$ (bottom) distributions. Data are shown by dot points with error bars, the signal by the dashed green curve, BKGI by the long-dashed pink curve, BKGII with $\omega$ intermediate state by the dotted cyan curve, BKGII with $\eta$ intermediate state by the dash-dotted red curve, and the total fit by the solid blue curve.}
\label{fig:projvisomega}
\end{center}
\end{figure}

Detailed MC studies indicate that the distributions of $M_{\ppp}^{\omega}$ and $M_{\ppp}^{\eta}$ are uncorrelated as no kinematic fit is performed. Therefore, in the 2D ML fit, a 2D PDF can be the direct product of the two one-dimensional (1D) PDFs for the two variables.
  Furthermore, MC studies validate that the 1D line shapes for the $\omega$ ($\eta$) meson are identical between the signal and the peaking background (BKGII). Consequently, the full 2D PDF used in the ML fit is constructed by
  \begin{align}
  \mathcal{F}&= N_{\rm sig}^{\rm data}\cdot F_{\rm sig}^{\omega}\cdot F_{\rm sig}^{\eta} + N_{\rm bkg}\cdot F_{\rm bkg}^{\omega}\cdot F_{\rm bkg}^{\eta} \nonumber  \\
  & + N_{\rm bkg}^{\omega}\cdot F_{\rm sig}^{\omega}\cdot F_{\rm bkg}^{\eta}+ N_{\rm bkg}^{\eta}\cdot F_{\rm bkg}^{\omega}\cdot F_{\rm sig}^{\eta},
  \label{likelihood}
  \end{align}
  where $N_{\rm sig}^{\rm data}$ is the number of signal events including the contribution from BKGIII, $N_{\rm bkg}$, $N_{\rm bkg}^{\omega}$ and $N_{\rm bkg}^{\eta}$ are the numbers of events for BKGI, and BKGII with $\omega$ and $\eta$ intermediate states, respectively.
  $F_{\rm sig}^{\omega}$ and $F_{\rm sig}^{\eta}$ are the PDFs of $\omega$ and $\eta$ signals in the $\pi^+\pi^-\pi^0$ invariant mass spectrum, respectively, described by the sum of two crystal ball (CB) functions~\cite{CB} with common mean and sigma values, but opposite side and different parameters for tails. $F_{\rm bkg}^{\omega}$ represents the non-$\omega$ component in the  $M_{\ppp}^{\omega}$ distribution, and is described by a second order Chebyshev polynomial function. $F_{\rm bkg}^{\eta}$ is the non-$\eta$ component in the $M_{\ppp}^{\eta}$ distribution,  and is represented by a reversed ARGUS function~\cite{argus}, defined as
\begin{equation}
F_{\rm bkg}^{\eta}(m) =  m \cdot (1-(X-m)^2/t^2)^a \cdot \exp(-b\cdot(1-(X-m)^2/t^2)),
\label{threshold}
\end{equation}
\noindent where $X$ is the sum of the lower and upper limits of the fit range, $a$ and $b$ are  constant coefficients, and $t$  is the upper limit of the fit range. All the parameters of Eq.~(\ref{likelihood}) are left free during the fit except the upper and lower limits of the fit range. 

The projections of the ML fit to the $M_{\ppp}^\omega$ and $M_{\ppp}^{\eta}$ distributions are shown in Fig.~\ref{fig:projvisomega}. The fit yields $N_{\rm sig}^{\rm data}= 32528  \pm 283$. After subtracting the contribution of BKGIII, the net number of signal events is $\Nsig= 31442 \pm 314$. By taking into account the signal yield $\Nsig$, the detection efficiency $6.2\%$ obtained from the corresponding MC sample, and the decay branching fractions of $\omegavis$ and $\etappp$ quoted in PDG~\cite{pdg}, the branching fraction of $\jpsiomeeta$ is measured to be larger by  $12.0\%$ with respect to its world average value quoted in the PDG~\cite{pdg},  but consistent within the uncertainty.

\subsection{The visible decay mode of  {\boldmath $\phivis$}}
\label{visibledecayphi}
\begin{figure}[htbp]
\begin{center}
 \includegraphics[width=0.45\textwidth]{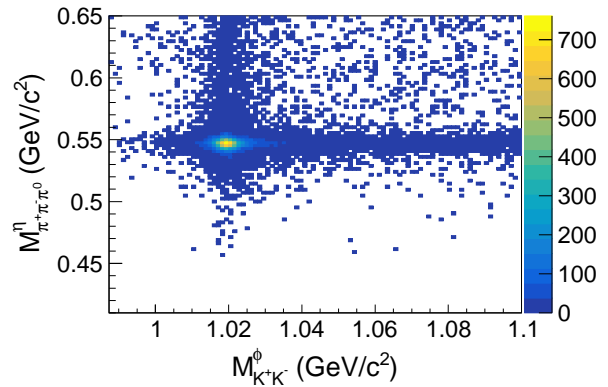}
 \caption{(Color online) Distribution of  $M_{\kk}^{\phi}$  versus $M_{\ppp}^{\eta}$  for data.}
\label{fig:2dplotphi}
\end{center}
\end{figure}

  For the candidate events of $\jpsiphieta$ with subsequent decays $\phivis$ and $\etappp$, the $\eta$ candidate is reconstructed with exactly same process as described in Sec.~\ref{visibledecayinv}, and the $\phi$ candidate is reconstructed by two additional oppositely charged tracks, which are assumed to be kaons without any PID requirement. 
  The total energy of the selected $\kk\ppp$ final state must satisfy  $\etot>2.95~\gev$. Similarly, the polar angle of the system recoiling against the $\eta$ candidate $\theta_{\rm recoil}$ is required to satisfy $|\cos\theta_{\rm recoil}| < 0.7$ to minimize the systematic uncertainty. The candidate events with invariant mass of $\kk$ in the range [0.987, 1.10]~$\gevcc$ are kept for further studies (Figure~\ref{fig:2dplotphi}). The remaining backgrounds are analogous to  BKGI, BKGII and BKGIII  in the $\jpsiomeeta$ visible decay obtained by replacing $\omega$ with $\phi$ signal, and the corresponding $\ppp$ with $\kk$. Similarly, the contributions of BKGI and BKGII are determined by a 2D ML fit, and the BKGIII of $238.6 \pm 26.0$ events, estimated with an exclusive MC sample of $\jpsiphieta$ with subsequent decays $\phivis$ and $\eta\to\gamma\pi^+\pi^-$ and normalized according to the branching fractions quoted in the PDG~\cite{pdg}, is subtracted from the  signal yield obtained from the 2D ML fit.

\begin{figure}[htbp]
  \begin{center}
 \includegraphics[width=0.4\textwidth]{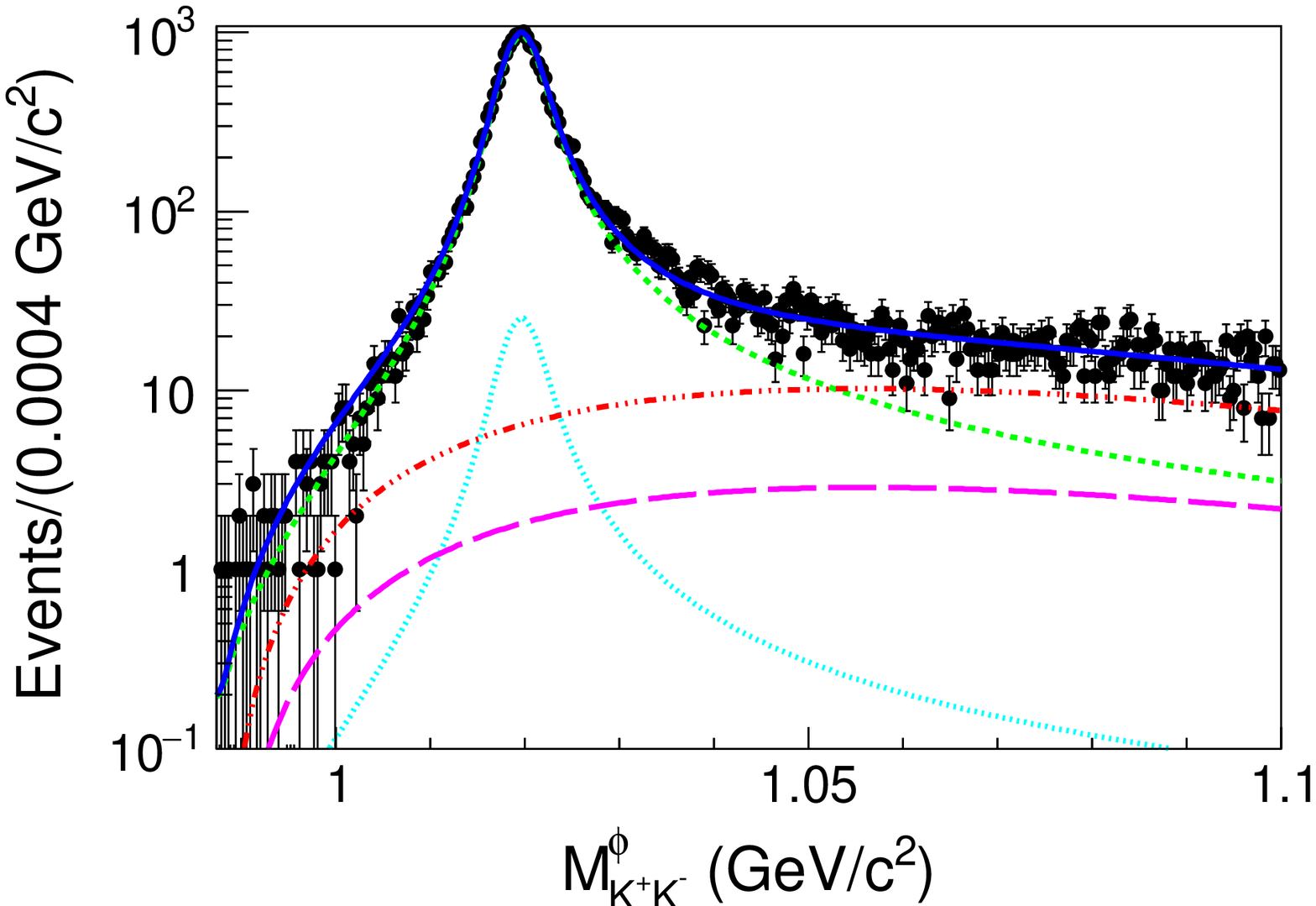}
\includegraphics[width=0.4\textwidth]{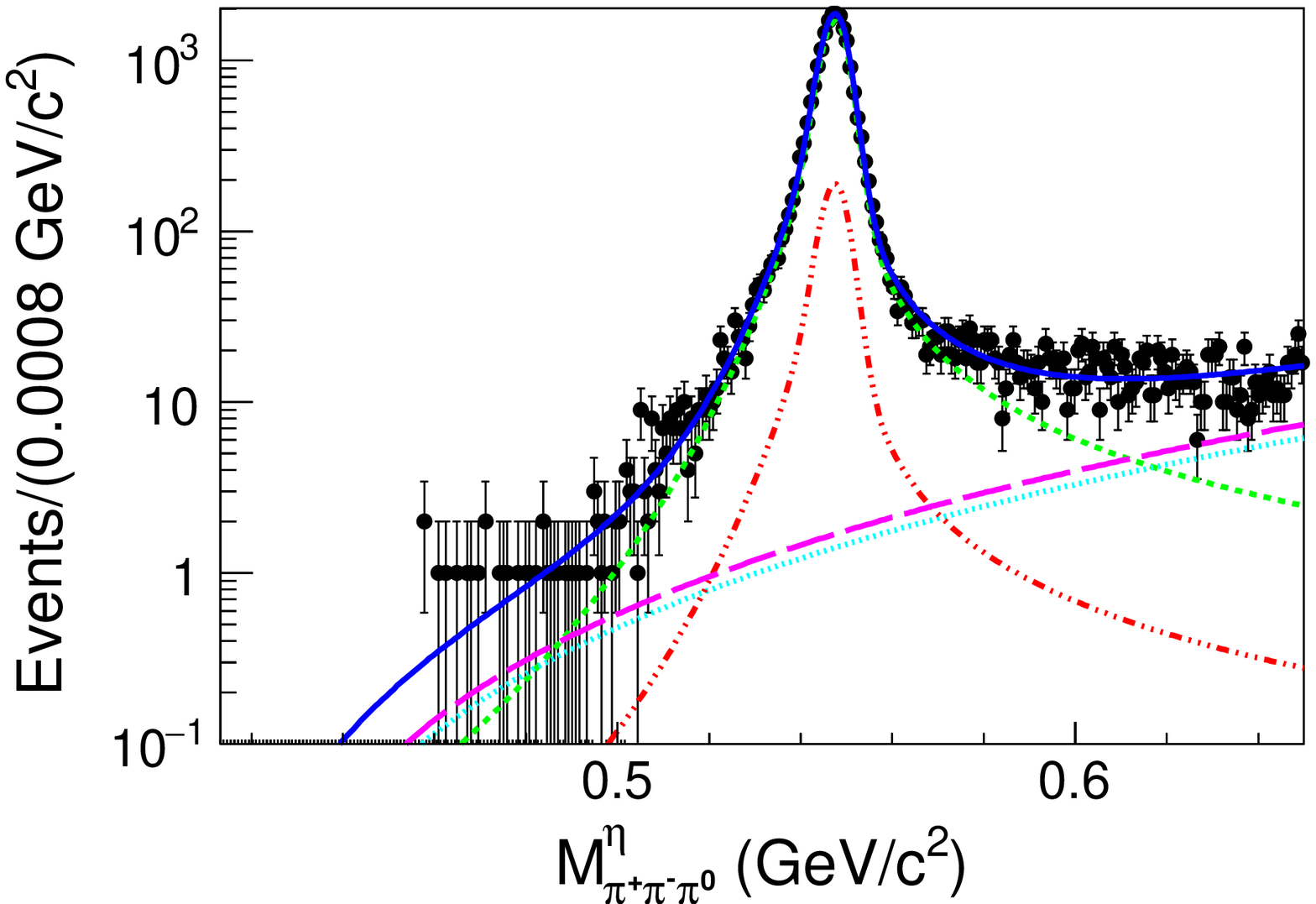}
 \caption{(Color online) Projections of the 2D fit to the  $M_{\kk}^{\phi}$ (top) and  $M_{\ppp}^{\eta}$ (bottom) distributions. The data are shown by the dots with error bars, the signal by the dashed green curve, BKGI by the long-dashed  pink curve, BKGII with $\phi$ intermediate state by the dotted cyan curve, BKGII with $\eta$ intermediate state by the dash-dotted red curve, and the total fit by the solid blue curve.}
\label{fig:projvis}
\end{center}
\end{figure}

  A similar 2D ML fit comprising $M_{\kk}^{\phi}$ and $M_{\ppp}^{\eta}$ is carried out to obtain the  signal yield. The parameterizations of the 1D PDF for $\eta$ and non-$\eta$ components in  $M_{\ppp}^{\eta}$  are the same as those used in the case of the $\jpsiomeeta$ visible decay. The 1D PDF for the $\phi$ signal in the $M_{\kk}^{\phi}$ distribution is described by a relativistic Breit-Wigner (BW)~\cite{pdg} function convolved with a Gaussian function representing  the mass resolution, where the mass and width of the $\phi$ signal are taken from the PDG~\cite{pdg}, and the parameters of the Gaussian function are left free in the fit.  For the non-$\phi$ components in  $M_{\kk}^{\phi}$, its 1D PDF is represented by a reversed ARGUS function as described in Sec.~\ref{visibledecayomega} by fixing the threshold parameter $t$ to the upper limit of the fit range.
  The ML fit yields $N_{\rm sig}^{\rm data} = 19534 \pm 186$, and the  $M_{\kk}^{\phi}$ and $M_{\ppp}^{\eta}$ projections are shown in Fig.~\ref{fig:projvis}.  The net number of signal events after subtracting the contribution of BKGIII from $N_{\rm sig}^{\rm data}$ is $\Nsig = 19295 \pm 188$. We compute the branching fraction of $\jpsiphieta$ by taking into account $\Nsig$, the detection efficiency 15.8\% from the corresponding MC sample, and the decay branching fractions of $\phivis$ and $\etappp$ from the PDG~\cite{pdg}. The measured branching fraction of $\jpsiphieta$ is also larger by $12.0\%$ over its world average value~\cite{pdg}, but consistent within the uncertainty.

\section{Systematic uncertainty}

  Since we measure the relative ratios of the branching fractions of invisible decay to that of corresponding visible decay, the systematic uncertainties associated with the  number of $\jpsi$ events, the reconstruction efficiency of $\etappp$, the requirement on $\cos\theta_{\rm recoil}$, the branching fractions of $\etappp$ and $\jpsi\to V\eta$  cancel. The remaining sources of  systematic uncertainties are associated with the fit procedure of the invisible and visible decays, the $\egamext$ requirement in the invisible decay, charged track reconstruction, trigger efficiency, photon detection and  the $\etot$ requirement for the visible decay and the branching fractions of $\omegavis$ and $\phivis$ decays. The details of the evaluation of individual uncertainties are described below and summarized in Table~\ref{systfinal}.

  The systematic uncertainty associated with the fit procedure in the invisible decays, which can reduce the significance of any observation, but does not scale with the reconstructed signal yields, is considered to be an additive systematic uncertainty. The other remaining sources of  systematic uncertainties, which don't not affect the significance of any observation, but scale with the number of reconstructed signal yield, are considered multiplicative systematic uncertainties.

  The systematic uncertainties associated with the tracking efficiency of kaon and pion are $1.0\%$ for each track, obtained by investigating the control samples of $\jpsi \to K_S^0 K^{\pm} \pi^{\mp}$  and $\jpsi \to \pi^+\pi^-p \overline{p}$, respectively. The systematic uncertainty of the photon reconstruction efficiency is less than 1.0\% per photon, investigated with a control sample sample of $\ee \to \gamma \mu^+\mu^-$ in which the four-momenta of two muons are used to obtain  the ISR photon momentum~\cite{vindyphot}. The uncertainty associated with the $\egamext$ requirement in the invisible decay is determined to be 1.1\% by comparing the corresponding detection efficiencies between data and MC simulation with a control sample of $\jpsi\to\ppp$. The systematic uncertainty due to $\pi^0$ reconstruction efficiency is determined to be $1.0\%$ using a control sample of $J/\psi \to p\overline{p}\pi^0$. The uncertainty associated with the $\etot$ requirement in the visible decay processes is explored with the relative efficiency with respect to an alternative requirement $\etot > 2.6~\gev$, where the signal loss is expected to be negligible. The relative differences in  efficiency between data and MC simulation, $2.1\%$  and $1.0\%$ for $\omega$ and $\phi$ visible decays, respectively, are considered as the uncertainties.

  The BESIII trigger system combines the information from the sub-detectors of EMC, MDC and TOF to select the events of interest for readout.   We study the trigger efficiency with a control sample  of $\jpsi\to\ppp$, and found the efficiency is almost $100\%$ for an event with two charged and two photons by considering the different kinematics of the final state. We assign $0.1\%$ as a systematic uncertainty related with the trigger efficiency.    

  The uncertainty associated with the fit procedure in invisible decays originates from the signal PDF, non-peaking background modelling and the fit bias. The uncertainty due to the signal PDF is estimated by an alternative fit with the sum of the two CB functions for the signal PDF, where the corresponding parameters of the CB functions are obtained by fitting the simulated MC samples and fixed in the fit. The uncertainty due to the non-peaking background shape is estimated by using an alternative PDF of a second order Chebyshev polynomial function in the fit. The relative changes in the results are taken as the uncertainties. A large number of pseudo-experiments with  fixed amount of signal, peaking and non-peaking background events expected from the data are generated to examine the bias of the fit procedure. The same fit procedure is repeated for each MC set, and the average shift of resultant signal yields can be taken as  systematic uncertainty, and is found to be negligible.

  The uncertainty related to the fit procedure for the reference decay is obtained in an analogous way. The uncertainty due to the fixed PDF parameters of the BW  in the $\phi \to K^+K^-$ decay is evaluated by varying each parameter within its statistical uncertainty while taking the correlations between the parameters into account. The uncertainty associated with the PDFs for the non-$\omega$ component on the $M^{\omega}_{\ppp}$ distribution is estimated by changing the order of the Chebyshev polynomial function.
  The uncertainties associated with the PDFs for  the non-$\eta$ component in  $M^{\eta}_{\ppp}$ and the non-$\phi$ component in  $M_{\kk}^{\phi}$  are estimated by modifying the formula of Eq.~(\ref{threshold}) as $F_{\rm bkg}^{\eta}(m) =   m^2/t \cdot (1-(X-m)^2/t^2)^a \cdot \exp(-b\cdot(1-(X-m)^2/t^2))$.  The largest relative change of the signal yields of individual alternative fits is considered as systematic uncertainty, and the total uncertainty associated with the fit procedure is the quadrature sum of the individual values.
  
  The uncertainties associated with the branching fractions of $\omega$ or $\phi$ visible decays are taken from the PDG~\cite{pdg}, and included only in the  results for the branching fractions. 

\begin{table}
\centering
\caption{Systematic uncertainties and their sources.}
\begin{tabular}{p{4.5cm} c c}
        \hline    \hline

 Source   & ~~$\omega$ decays~~ &  ~~$\phi$ decays~~  \\
\hline \hline
\multicolumn{3}{c}{Additive systematic uncertainties (events) } \\
\hline
\hline
Fixed PDFs                  & 0.1   & 0.1      \\
 Background modelling       & 1.6   & 1.0   \\
\hline
Total                       & 1.6   & 1.0  \\
\hline
\hline
\multicolumn{3}{c}{Multiplicative systematic uncertainties ($\%$)}     \\
\hline
\hline

Charged tracks reconstruction                       & 2.0   & 2.0    \\
Photon detection                                    & 2.0   & ---     \\
$\egamext$ requirement                              & 1.1   & 1.1    \\
$\pi^0$ reconstrunction                             & 1.0   & ---    \\
$\etot$ requirement                                 & 2.1   & 1.0    \\
Fit parameters (visible decays)                     & 0.3   & negl.  \\ 
$\mathcal{B}(\omegavis/\phivis)$                    & 0.8   & 1.0    \\ 
$N_{\rm sig}^{\rm visible}$ uncertainty                  & 1.0   & 1.0    \\
Trigger efficiency                                  & 0.1   & 0.1    \\ \hline
Total                                               & 4.0   & 2.9    \\ \hline \hline
\end{tabular}
\label{systfinal}
\end{table}

\section{results}
No obvious signal for $\omega$ and $\phi$ invisible decays is observed.  We compute the upper limits on the ratio of branching fractions of the invisible decay to that of  the corresponding visible decay, $\Romega$ and $\Rphi$, at the $90\%$ confidence level (C.L.) using the Bayesian approach~\cite{pdg}, individually. The branching fraction ratios of $\Romega$ and $\Rphi$ are calculated using the formula of Eq.~(\ref{bfrat}) after incorporating obtained signal yields and the corresponding detection efficiencies for the visible and invisible decays as presented above. The systematic uncertainty is included by convolving the likelihood  versus the branching fraction ratio curve with a Gaussian function with a width equal to the systematic uncertainty. The upper limits on the branching fraction ratios  are measured to be  $ \Romega < \numRomega$ and  $\Rphi < \numRphi $ for $\omega$ and $\phi$ mesons, respectively, at the 90\% C.L.\ after integrating their likelihood versus branching fraction ratio curves from zero to $90\%$ of the total curve. By using the branching fractions of $\omegavis$ and $\phivis$ quoted in the PDG~\cite{pdg}, the  upper limits on the invisible decay branching fractions at the 90\% C.L.\ are calculated to be $\Bomega< \numBomega$ and $\Bphi<\numBphi$, individually.

\section{Summary}
Using a data sample of $\numjpsi$ $\jpsi$ events collected by the BESIII experiment at the BEPCII collider, a search for the invisible decays of $\omega$ and $\phi$ mesons in $\jpsi\to V\eta$ decays is performed for the first time. We find no significant signal for these invisible decays and set $90\%$ C.L. upper limits on the ratio of branching fractions of invisible decays to that of the corresponding visible decays to be $\Romega < \numRomega$ and $\Rphi  < \numRphi$, respectively. The upper limits on the branching fractions $\Bomega$ and $\Bphi$ are also determined to be less than $\numBomega$ and $\numBphi$, respectively, at the $90\%$ C.L.\ by using $\mathcal{B}(\omega \to \pi^+\pi^-\pi^0)$ and $\mathcal{B}(\phi \to K^+K^-)$ from the PDG~\cite{pdg}.  These results can provide a complementary information to study the nature of dark matter and constrain the parameters of phenomenological models~\cite{Nicolas,invisibleuds}.

\acknowledgments

The BESIII collaboration thanks the staff of BEPCII, the IHEP computing center and the supercomputing center of USTC for their strong support. This work is supported in part by National Key Basic Research Program of China under Contract No. 2015CB856700; National Natural Science Foundation of China (NSFC) under Contracts Nos. 11335008, 11375170, 11425524, 11475164, 11475169, 11605196, 11605198, 11625523, 11635010, 11705192, 11735014; the Chinese Academy of Sciences (CAS) Large-Scale Scientific Facility Program; the CAS Center for Excellence in Particle Physics (CCEPP); Joint Large-Scale Scientific Facility Funds of the NSFC and CAS under Contracts Nos. U1532102, U1532257, U1532258,  U1632107, U1732263; CAS Key Research Program of Frontier Sciences under Contracts Nos. QYZDJ-SSW-SLH003, QYZDJ-SSW-SLH040; 100 Talents Program of CAS; INPAC and Shanghai Key Laboratory for Particle Physics and Cosmology; German Research Foundation DFG under Contracts Nos. Collaborative Research Center CRC 1044, FOR 2359; Istituto Nazionale di Fisica Nucleare, Italy; Koninklijke Nederlandse Akademie van Wetenschappen (KNAW) under Contract No. 530-4CDP03; Ministry of Development of Turkey under Contract No. DPT2006K-120470; National Science and Technology fund; The Swedish Research Council; U. S. Department of Energy under Contracts Nos. DE-FG02-05ER41374, DE-SC-0010118, DE-SC-0010504, DE-SC-0012069; University of Groningen (RuG) and the Helmholtzzentrum fuer Schwerionenforschung GmbH (GSI), Darmstadt.

\end{document}